\def\year{2017}\relax
\documentclass[letterpaper]{IEEEtran}

\IEEEoverridecommandlockouts

\usepackage{hyperref}
\usepackage{times}
\usepackage{helvet}
\usepackage{courier}
\usepackage[inline]{enumitem}
\usepackage{graphicx}
\usepackage{amsmath}
\usepackage{multirow}

\usepackage{array}
\newcolumntype{P}[1]{>{\centering\arraybackslash\hspace{0pt}}p{#1}}

\begin{document}

\title{Identifying the social signals that drive online discussions: A case study of Reddit communities}

\author{Benjamin D. Horne, Sibel Adal{\i}, and Sujoy Sikdar\\
Rensselaer Polytechnic Institute\\
110 8th Street, Troy, New York, USA\\
\{horneb, adalis, sikdas\}@rpi.edu
}

\IEEEpubid{\makebox[\columnwidth]{\hfill ~\copyright~2017 IEEE}
\hspace{\columnsep}\makebox[\columnwidth]{
IEEE International Conference of Computer Communications and Networks}}
\maketitle

\begin{abstract}
Increasingly people form opinions based on information they consume on
online social media. As a result, it is crucial to understand what
type of content attracts people's attention on social media and drive
discussions. In this paper we focus on online discussions. Can we
predict which comments and what content gets the highest attention in
an online discussion?  How does this content differ from community to
community? To accomplish this, we undertake a unique study of Reddit
involving a large sample comments from 11 popular subreddits with
different properties. We introduce a large number of sentiment,
relevance, content analysis features including some novel features
customized to reddit. Through a comparative analysis of the chosen
subreddits, we show that our models are correctly able to retrieve top
replies under a post with great precision. In addition, we explain our
findings with a detailed analysis of what distinguishes high scoring
posts in different communities that differ along the dimensions of the
specificity of topic and style, audience and level of moderation.
\end{abstract}

\begin{figure*}[ht]
\centering
\includegraphics[width=0.9\linewidth]{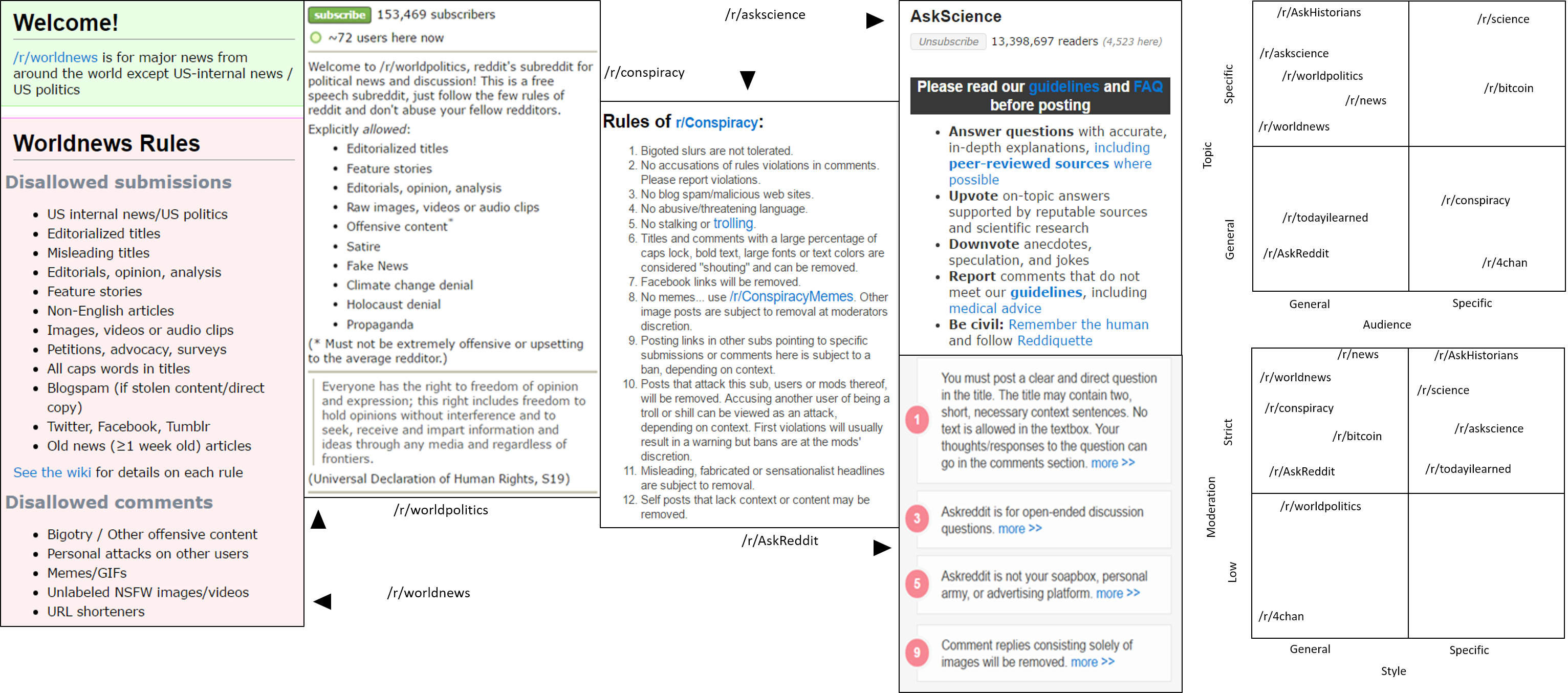}
\begin{tabular}{P{0.7\linewidth}P{0.3\linewidth}}
(a) Example subreddit rules & (b) Dimensions of subreddits \\
\end{tabular}
\caption{Rules \& dimensions: Wide variation in rules can be seen between r/worldnews \& r/worldpolitics in part a and in the moderation dimension of part b.}\label{collage}
\end{figure*}

\section{Introduction}
In this paper, we study the following problem: What type of comments drive discussions on social media? First, we examine whether it is possible to predict which comments receive positive attention. In conjunction, we ask the following related questions: If prediction is possible, what features are useful in prediction? Secondly, what are the distinguishing features of comments that receive high attention in each community, and how do these differ from one community to another?


Increasingly, people form opinions based on information they consume on
online social media, where massive amounts of information are
filtered and prioritized through different communities. As a result,
social media sites are often targets of campaigns for dissemination of
information as well as misinformation~\cite{Bessi:2015hr}~\cite{Anonymous:9zcY0p44}. 
These campaigns can employ sophisticated techniques to hijack discussions by posting content with a specific point of view within posts and discussions in order to attract attention and steer the discussion or influence how users interpret the original content~\cite{maddock2017}.
It is often observed that in our information saturated world, user attention is
one of the most valuable commodities.  Hence, it is crucial to
understand which type of content receives high attention. 
We consider this as the first step in the study of information dissemination in online discussions and in building tools for information processing. 

To address the central problem of this paper, we study a large dataset
of comments from many different communities on reddit. Reddit is one
of the most popular platforms for news sharing and discussion, ranking
\#4th most visited site in US and \#16 in the
world. Reddit claims to be the front page of
the internet, achieving its stated purpose by allowing users to {\em
  post} news, questions, and other information in the form of text,
images and links to external websites. Users often engage with the
posts by getting involved in or reading discussions consisting of
comments made by other users in the community. Discussions are a vital
and valuable feature of Reddit. Posts often generate lengthy and
vibrant discussions, and comments that help users analyze and engage
with the content, through the different perspectives and
interpretations provided by members of the community.

Voting is the main mechanism reddit provides its users to affect the
ranking and the visibility of posts and comments. Every post or
comment on reddit is assigned a {\em score} based on the votes it
receives. An {\em upvote} increases the score by one and a {\em
  downvote} decreases it. Posts and comments are sorted and presented
to users (loosely) in order of the score they receive. While reddits'
algorithm slightly obfuscates the ordering to prevent users from
gaming the mechanism, the score is the primary and most significant
factor in ordering posts and comments made within a small time period
and is directly correlated with the votes. Voting allows users to
steer the discussion and drive the most relevant, interesting, and
insightful comments to prominence in the discussion.

\IEEEpubidadjcol

It is undeniable that reddit has its own norms and culture, organized
around its communities called subreddits. Subreddits differ from each
other in many different ways, especially in four specific dimensions:
topic, audience, moderation and style (see Figure~\ref{collage}). Some
subreddits are topic specific (\texttt{r/AskHistorians} or
\texttt{r/Bitcoin}), while others are from a general topic
(\texttt{r/AskReddit}).  Subreddits can differ in whether they target
a specialized audience or not as in the case of \texttt{r/Bitcoin} for
experts on this topic. Some subreddits have a very specific style for
posting questions and comments such as in \texttt{r/todayilearned}
that targets submissions that are verifiable facts. While all
subreddits have rules regarding what types of content is allowable in
that community (see Figure~{collage}), the specificity of the rules
and the level of moderation differ greatly from subreddit to
subreddit. Even when the written standards are similar, reddit
communities attract users with different interests and discussions of
different nature. One expects that this results in other {\em
  unwritten} standards of quality that can only be inferred from the
readers' votes.

Given the very large user base of reddit and the diversity of
subreddits along these four dimensions, it is not necessarily clear
that high scoring content is predictable. In fact, a recent study
reports fairly low accuracy results~\cite{jaech2015talking}. Previous
work does not make it clear to which degree the prediction accuracy is
impacted by the choice of features, communities or the learning
method. We study this problem in detail and make the following
contributions:

\begin{itemize}
\item We undertake a study featuring 11 popular subreddits that differ
  across the four dimensions discussed above. 
We sample a large number of posts analyze the content of comments under them and their relationship to  scores.

\item We analyse a comprehensive set of features, from previous work
  for predicting expertise, news engagement and readability as well as
  novel features geared towards the reddit culture, such as the
  self-referential nature of discussions.

\item We train machine learning models using a combination of time,
  sentiment, relevance and content analysis features. Our models
  significantly outperform the state of the art and perform well
  across a range of subreddits, irrespective of topic, moderation,
  audience and style, consistently ranking the top comments by score
  with high precision. We find that sentiment based features are more
  useful than other categories.



\item We perform a post-hoc analysis to find significant features that
  distinguish between high and low scoring comments. Many features are
  significant in many communities including our novel features. Some
  features are consistently positively or negatively correlated across
  communities, while others may flip sign between communities. Most
  notably the relationship to time of comments shows a more complex
  picture than previously reported in the literature.

\item We include a detailed discussion of the similarities and
  differences between communities based on this post-hoc analysis. We
  show that audience, specificity of topic and style matter greatly in
  understanding which features are prominent. Surprisingly, we find
  that subreddits with very different levels of moderation may show
  very similar behavior.


\item We also study the impact of users' attributes and show
  that comments by high scoring and highly active users do not
  necessarily end up on top. However, comments by users with flairs
  often end up on top, even when these flairs are self-assigned. We
  speculate that flairs act as an easy to evaluate heuristic signaling
  expertise.

\end{itemize}

\section{Related Work}
Most related to our work is work by Jaech et
al~\cite{jaech2015talking}, in which the authors explore language's
impact on reddit discussions. The authors use a set of many complex
natural language features to rank comment threads in 6 different
subreddits using a SVM ranking algorithm. The ranking results achieve
relatively low predictive power, only attaining an average of 26.6\%
precision in retrieving the top 1 comment correctly. However, these
results show that feature importance can change across community
types. Along with this, the authors study the relative impact of
``high karma'' users on discussions by computing the percentage of
discussions where the top comment is made by the top h-index user,
where h-index is the the number of comments in each user's history
that have a score (karma) greater than or equal to $k$.  This brief
user analysis concluded that high h-index users have little impact on
the popularity of a comment. For comparison, our work will have some
overlaps with this work. Specifically, we will use 2 of the same
communities, have some feature overlap, and use the h-index
calculation in our user analysis. However, we incorporate many novel
features, study a much larger data set with many more communities,
achieve a far greater accuracy and incorporate a unique post-hoc
analysis showing striking differences across different communities.

In addition to Jaech et al~\cite{jaech2015talking}, reddit has been
well-studied from many different perspectives. Dansih et al. show that
a comment's timing relative to the post matter in eliciting responses
in the community
\texttt{r/IAmA}~\cite{dahiya2016discovering}. Lakkaraju et
al/~\cite{lakkaraju2013s} and Tran et
al.~\cite{tran2016characterizing} show that reddit post titles and
timing are important factors in the popularity of a post. However,
this popularity can be delayed. In 2013, Gilbert determined that
roughly 52\% of popular reddit posts are unnoticed when first
submitted. Further, the popularity and engagement of reddit posts can
be reasonably determined by many
factors~\cite{gilbert2013widespread}. Althoff et al. illustrate that
temporal, social, and language features can play a role in successful
requests in a study of altruistic requests in the reddit community
\texttt{r/randomactsofpizza}~\cite{althoff2014ask}. More recently,
Hessel, Lee, and Mimno show that visual and text features are
important in image-based post popularity prediction. More over,
Hessel, Lee, and Mimno show that user-based features do not predict
popularity as well~\cite{hessel2017cats}. While all of these studies
are exploring the posts on reddit rather than the comments as we do in
this paper, they demonstrate the many perplexities in both the messages
and messengers on reddit.

Also related to this problem is work on general information
popularity, news engagement, expert finding, and information
credibility.  Sikdar et al.~\cite{sikdar2013understanding} develop
models to predict credibility of messages on Twitter using several
user and natural language features. This work shows that crowdsourced
endorsements like upvoting contribute to predicting information
credibility. Note that credibility of content does not necessarily
imply its correctness, as one of the subreddits we choose explicitly
allows conspiracy theories. A recent 2017 study explores the impact of
reddit on news popularity in the community
\texttt{r/worldnews}~\cite{horne2017impact} and finds that well-known
news popularity metrics are able to accurately predict the popularity
of a news article on reddit. In essence, news behavior on this
subreddit resembles news popularity in general. This work also shows
that users tend to the change the news titles to be more positive and
more analytical despite news being more negative
overall~\cite{reis2015breaking}. When users change the titles of a
news article, the article tends to become more popular. Similarly,
another study predicts the popularity of news using sentiment and
language features, showing that sentiment features are important in
popularity prediction~\cite{keneshloo2016predicting}. In 2016, Horne
et al.~\cite{horne2016expertise} studied automatic discovery of
experts on Twitter using simple language and meta heuristics. They
found that experts tend to be more active on Twitter than their
friends and that experts use simpler language than their friends, but
more technical langauge than the users they mention. We will borrow
features from many of these studies in our analysis of reddit comments
as credibility and expertise are part of popularity of a comment.

\begin{table*}[thb!]
\centering
\begin{minipage}[t]{3.2in}
\begin{tabular}{p{0.6in}p{2.2in}} 
Abbr. & Description \\ \hline
h-index &  number of comments in each user’s local history
that have score $\geq$ h\\
activity & \# of comments + \# of posts made by user\\
flair & has flair or not, displayed with author’s username, assigned by a moderator or self depending on subreddit	\\ \hline 
     \multicolumn{2}{c}{(a) User Features}  \\\\
     \hline
time\_diff & difference between post and comment time \\ \hline 
     \multicolumn{2}{c}{(b) Time Features}  \\\\
     \hline
  	vad\_neg & negative sentiment score Vader-Sentiment\\
  	vad\_pos & positive sentiment score Vader-Sentiment\\
  	vad\_neu & neutral sentiment score Vader-Sentiment\\
  	vad\_comp & composite sentiment score Vader-Sentiment\\
  	psubj & probability of subjectivity using a learned Naive Bayes classifier\\
  	pobj & probability of objectivity using a learned Naive Bayes classifier\\
  	subjcat & binary category of objective or subjective\\
  	posemo & number of positive emotion words\\
  	negemo & number of negative emotion words\\
  	tone & number of emotional tone words\\
  	affect & number of emotion words (anger, sad, etc.)\\
	analytic & number of analytic words\\  	
  	insight & number of insightful words\\
  	authentic & number of authentic words \\
  	clout & number of clout words \\
  	tentative & number of tentative words \\
  	certain & number of certainty words \\
  	affil & number of affiliation words \\
  	focuspresent & number of present tense words\\
  	focusfuture & number of future tense words\\
  	focuspast & number of past tense words\\
  	\hline 
\multicolumn{2}{c}{(c) Sentiment Features (including subjectivity)} 
\end{tabular}
  \end{minipage} 
\begin{minipage}[t]{3.2in}
\begin{tabular}{p{0.6in}p{2.4in}}  
Abbr. & Description \\ \hline
relevance & cosine similarity between post vector and comment vector\\   	\hline 
\multicolumn{2}{c}{(d) Relevance Features} \\ \\ \hline
self\_fluency & avg. frequency of most common words in subreddit corpus\\
  coca\_fluency & avg. frequency of least common 3 words using coca corpus \\
  WC & word count\\
  WPS  & words per sentence\\
  GI & Gunning Fog Grade Readability Index\\
  SMOG & SMOG Readability Index\\
  FKE & Flesh-Kincaid Grade Readability Index\\
   ttr & Type-Token Ratio (lexical diversity) \\
   conj & number of conjunctions \\
   adverb & number of adverbs \\
   auxverb & number of auxiliary verbs \\
   pronoun & number of pronouns \\
   ppron & number of personal pronouns\\
    i & number of I pronouns\\
  we & number of we pronouns \\
  you & number of you pronouns\\
  shehe & number of she or he pronouns\\
  quant & number of quantifying words\\
  swear & number of swear words \\
  netspeak & number of online slang terms (lol, brb)\\
  interrog & number of interrogatives (how, what, why)\\
  per\_stop & percent of stop words (the, is, on)\\
  AllPunc & number of punctuation\\
  quotes & number of quotes\\
  function & number of function words \\
  word\_len & average word length \\
  \hline
  \multicolumn{2}{c}{(e) Content Features} \\ \\
\end{tabular}
\end{minipage}
\caption{\label{tbl:features} Different features used in our study}
\end{table*}

\section{Features}
To explore reddit discussions, we compute features on each comment in
our data set. These features can be categorized into five
categories: \begin{enumerate*}[label=(\arabic*)] \item sentiment
   \item content \item relevance \item user, and \item
  time \end{enumerate*}. A short description of our feature sets can
be found in Table~\ref{tbl:features}.

\paragraph{Sentiment and Subjectivity}  To compute sentiment features, we
utilize the tool Vader-Sentiment~\cite{hutto2014vader} which is a
lexicon and rule-based sentiment analysis tool proposed in 2014. We
choose this tool as it is specifically built for ``sentiment expressed
on social media~\cite{hutto2014vader}'' and has been shown to work well
on reddit and news
data~\cite{horne2017impact}~\cite{keneshloo2016predicting}. It
provides 4 scores: negative, positive, neutral, and composite, where
composite can be thought of as the overall sentiment in a text. We
will include all 4 scores as features in our sentiment/subjectivity
model.

Next, we utilize the Linguistic Inquiry and Word Count
(LIWC) tool~\cite{tausczik2010psychological} for a mix of features for
emotion and perceived objectivity of a comment. The emotion features
include: positive and negative emotion, emotional tone, and
affect. Other features from LIWC included in our
sentiment/subjectivity model are: analytic, insight, authentic,
clout, tentative, certainty, affiliation, present tense, future tense,
and past tense. Despite LIWC computing these features using simple
words counts, they have been shown to work well in a variety of
settings~\cite{wu2011does}.

Further, we include three features that directly measure the
probability a comment is subjective or objective, computed by training
a Niave Bayes classifier on 10K labeled subjective and objective
sentences from Pang and Lee~\cite{pang2004sentimental}. The classifier
achieves a 92\% 5-fold cross-validation accuracy and has been shown
useful in predicting news popularity~\cite{horne2017impact}.

\paragraph{Content Structure} To analyse content structure, we take other
word count features from LIWC such as parts of speech features
(similar to what a POS tagger would provide), punctuation, and word
counts for swear words and online slang. In addition, we capture the
readability and clarity of a comment using three metrics: Gunning Fog,
SMOG, and Flesch Kincaid, and the lexical diversity metric (Type-Token
Ratio)~\cite{horne2017impact}.

Next, we compute ``fluency'' features based on the (log) frequency of
words in a given corpus, capturing the relative rarity or commonality
of a piece of text. These features can mean several things depending
on the corpus used. Commonness of a word is in general is based on the
Corpus of Contemporary American English (COCA)~\cite{coca}.
It has been shown that humans tend to believe information that is more
familar, even if that information is
false~\cite{lewandowsky2012misinformation}. To capture how well a
comment fits into a given community style, we compute fluency on the
corpus of each community. This localized fluency captures the
well-known ``self-referential'' behavior of reddit and it's independent
communities~\cite{singer2014evolution}. Some communities may have a
very specific ``insider'' language, while others will not.

\paragraph{Relevance} To measure how much new information is added to
a comment, we compute the similarity between the text of a comment and
the post it is under as a notion of relevance of the comment to the
post. To compute this feature, we first vectorize each word using
word2vec~\cite{word2vec} trained on COCA. Once all words are in vector
format, we can compute centroids weighted by the inverse of the log
word frequency in the text. We do this for the 5 rarest words in the
post and the comment. Finally, we compute the cosine similarity
between the post vector and the comment vector. High similarity
suggests little new information was added to the discussion, while low
similarity suggests the opposite.

\paragraph{User} To study the influence of users on comment scores, we will
use three simple features. Local h-index is the number of comments
with score greater than or equal to $h$ within a given community. This
index is widely used to measure the scientific output of a
researcher and reddit karma
in~\cite{jaech2015talking}. This metric should capture a user's
reputation in a community better than any central measure based on a
user's historic comment scores.  Local activity of a user is defined
as the number of comments plus the number of posts a user makes within
the community. Finally, flairs are visual badges displayed next to a
user's screen name. Flairs are typically used to show a user's area of
expertise and are given out by the moderators through a strict
application process. However, some communities set these flairs up to
be arbitrary user-selected tokens. These two types of flairs mean very
different things as one is for expertise and the other simply for
community involvement.

\paragraph{Time}  To capture the timing of a comment, we will compute the
difference between the post and comment submission
times. Ranking using this time difference will be used as a baseline
model.

\begin{table}[h!]
\caption{Subreddits used in study ({\em Please note that some subreddits
    are NSFW and may contain offensive material. We highly recommend
    you use private browsing when visiting these
    subreddits.})}\label{subs}
\centering
\begin{tabular}{|c|c|c|p{0.5in}|p{0.25in}|}
\hline
\textbf{subreddit} & \textbf{\# posts} & \textbf{\# cmnts} & \textbf{\% users w/ flairs} & \textbf{\# flairs}\\
\hline
r/4chan & 10225 & 89080 & 5.6 & 133\\
r/AskHistorians  & 10381 & 21926 & 3.2 &304\\
r/AskReddit  & 10421 & 128487 & 0.0 &0\\
r/askscience  & 10404 & 23302 & 3.9 &637\\
r/Bitcoin  &10468 & 52639 & 0.0 &0\\
r/conspiracy  &10471 & 45103 & 0.0 &0\\
r/news  & 10583 & 46722 & 0.0  &0\\
r/science  & 10533  &  59307 & 1.7 &237\\
r/todayilearned  & 10453  &  102029 & 0.0 &0\\
r/worldnews  & 10388  &  78120 & 0.0 &0\\
r/worldpolitics  & 8057  &  24054 & 0.0 &0\\
\hline
\textbf{Total} & \textbf{112K} & \textbf{582K} &  &\textbf{1.3K}\\
\hline
\end{tabular}
\end{table}

\section{Data Sets}
To understand how discussion changes across communities, we gather
comment threads from 11 different subreddits during a 6 month period
in 2013. Once the comment threads are extracted, we randomly sample
10K comment threads from each subreddit. This data is extracted from
Tan and Lee’s reddit post data set~\cite{tan2015all} and Hessel et
al.’s full comment tree extension to that reddit
dataset~\cite{hessel2016science}, which contains 5692 subreddits, 88M
posts, and 887.5M comments between 2006 and 2014. The statistics on
our final extracted data sets can be found in Table~\ref{subs}.

\paragraph{Communities}
To understand the variation in noise and signal in online discussions,
we collect 11 communities with respect to 4 dimensions: topic,
audience, style, and moderation.  We explore communities that differ
widely in moderation (\texttt{r/worldnews} and
\texttt{r/worldpolitics}), communities based on expertise
(\texttt{r/science} and \texttt{r/askscience}), communities based on
news discussion (\texttt{r/news} and \texttt{r/worldnews}),
communities that have large general audiences, (\texttt{r/AskReddit}
and \texttt{r/todayilearned}), and communities that have smaller niche
audiences (\texttt{r/Bitcoin} and \texttt{r/conspiracy}). In addition,
we study \texttt{r/4chan}, a well-known ``troll'' and hate community
that reaches a very specific audience using very little moderation.

Figure~\ref{collage}b shows where each community in our study falls
with respect these four dimensions. In Figure~\ref{collage}a we
provide example subreddit rules.

\section{Methodology}
To understand the voting behavior of different communities on reddit
and to recover and uncover the communities' explicitly stated and
hidden quality standards, we use the following methodology: (1) Learn
a model to predict the score of a comment. (2) Evaluate learned models
using learning to rank metrics. (3) Perform post-hoc analysis.

\subsection{Learning to rank comments by score}
We first describe the experimental setup. As described earlier, each
subreddit consists of posts, under which users make comments. As a
basic preprocessing step, we remove all posts that have fewer than 5
comments under them as the frequency distribution of the number of
comments under a post is heavily skewed towards posts with just 1 or 2
comments. Including them in the dataset would heavily influence the
learning to rank metrics such as the average precision and render them
meaningless.  Each dataset corresponds to a subreddit, and consists of
comments, each described by a feature vector as well as information
about the user who made the comment and the post under which the
comment was made. In each subreddit, we pick 80\% of the posts
uniformly at random and use the comments under these posts to form the
training set. The remaining comments form the test set.

\subsubsection{Learning to predict score}
We learn a regression model on the comments in the training set where
each comment is described by a feature vector (see
Table~\ref{tbl:features}) and the predicted variable is the score of
the comment. In order to allow for easy introspection of the learned
models, and in light of the non-linearity of some of our features, we
chose to train simple linear models using the Python scikit-learn
library~\cite{pedregosa2011scikit}. We report results obtained from a
model learned using ridge regression with regularization, where the
regularization parameter is learned using 10-fold cross validation and
the optimization objective is to minimize the $L_2$-norm between the
predicted scores and the real scores from reddit data since it
performed the best overall.

The learned model is used to predict the scores of the comments in the
test set. We then rank the comments under each post in the test set
according to their predicted score. We measure the performance of our
models by comparing the predicted rankings versus the rankings
according to their true scores on reddit.

\subsubsection{Learning to rank metrics}
We evaluate the performance on the test set by the following metrics
from the learning to rank literature~\cite{Manning:2008:IIR:1394399}:
\begin{enumerate}
\item Average Precision @ k: The percentage of the posts ranked among
  the top $k$ as predicted by the learned model that are also among
  the top $k$ posts by true scores, averaged over all posts. 

\item Kendall-tau distance (KT-distance) @ k: Kendall-tau
  distance~\cite{kendall1938new} between the relative ranking of the
  top $k$ posts according to their true scores versus the relative
  ranking of the same $k$ posts by their predicted scores.
\end{enumerate}

 We report the precision for $k=1,3,5,10$ and KT-distance for
 $k=5,10,20$. KT-distance is a secondary feature, especially useful
 for posts that have a significantly large number of comments, giving
 us a complete picture together with precision.  If we achieve high
 precision for posts with large number of comments and the Kendall-tau
 distance is low at some value of $k$, it means
 that: \begin{enumerate*}\item comments predicted to be among the
   top $k$ were truly among the top $k$ by their true scores,
   and \item the relative positions of the true top $k$ comments are
   maintained in the predicted ranking.\end{enumerate*} 
To summarize, a good model displays the following qualities:
\begin{itemize}
\item High average precision at low values of $k$.
\item Low KT-distance for high values of $k$.
\item KT-distance grows sub-linearly with $k$.
\item At high values of $k$, high average precision and low KT-distance.
\end{itemize}

Good performance of learned models as measured by learning to rank
metrics validates the predictive and descriptive power of the
features. However, these models can still be hard to interpret. In
order to gain greater insight into the voting behavior of reddits
users in each community, we perform additional post-hoc analysis.

\subsection{Post-hoc Feature Analysis}
Our goal is to understand how each feature affects the score obtained
by a comment. We start by dividing the data into two classes: low
score comments (whose score are below the 50th percentile) and high
score comments (whose scores are above the 90th percentile). How does
the distribution of each feature affect whether a comment receives a
low score or a high score? Since the features are not usually
distributed normally, we use the Kolmogorov-Smirnov (KS) statistic as
a robust measure of the effect size, which is independent of the
distributions of the feature, and is sensitive to differences in the
middle of the distribution which is of particular interest for this
work. We then capture the top 15 features by effect size from each
subreddit that were significant (with a p-value less than 0.05).  We
also capture the difference between the mean of the distributions of
the feature values corresponding to the two classes to understand
whether high scoring comments are attributed with higher or lower
values of the feature.

\section{Results and Discussion}
In this section, we present results that answer the questions we set
out to address in the introduction.

\subsection{Yes, we can predict how comments are ranked.}
The performance of our learned models are summarized in
Table~\ref{tbl:ltr} which show that we can indeed predict ranking of
comments with high precision. This is consistent across
subreddits and the dimensions of style, moderation, subject and
target audience. We achieve high average precision at all values of
$k$ including at $k=1$. Moreover, the Kendall-Tau distance at $k$
grows roughly linearly with the value of $k$ (as opposed to growing
exponentially). Our models significantly outperform the state of the
art model~\cite{jaech2015talking}. Significantly, we achieved a
significantly higher average precision at 1 result of 0.412 and 0.671
in the \texttt{r/askscience} and \texttt{r/worldnews} subreddits
respectively (a 2 to 3 times improvement).

Since timeliness (represented by the feature time\_dif) of a comment
is widely cited as being a good predictor of the score of a comment
(and in other literature of a post), we used a model trained using
only timeliness as a feature as the baseline. Contrary to this widely
held belief, we find that timeliness alone does not guarantee a high
score. We also included time\_dif as a feature in our sentiment,
relevance and content models to measure the incremental improvement in
performance by using these features.  We found that relevance and
sentiment alone are both highly predictive of the score of a
comment. Sentiment held the highest predictive performance across
subreddits. Unfortunately, but not surprisingly, the models performed
poorly on the \texttt{r/AskReddit} dataset. This is likely because
\texttt{r/AskReddit} is simply too diverse in terms of its topic and
is aimed at a very general audience. It also somewhat loosely
moderated. Surprisingly, prediction was possible for other loosely
moderated subreddits such as \texttt{r/worldpolitics} and
\texttt{r/4chan}.

\begin{table*}[htp!]
\centering
\caption{\label{tbl:ltr}Evaluation of models}
\begin{tabular}{|c|c|c|c|c|c|c|c|c|} \hline
\multirow{2}{*}{Dataset} & \multirow{2}{*}{Model} & \multicolumn{4}{|c|}{Precision @ $k$} & \multicolumn{3}{|c|}{KT-distance @ $k$} \\ \cline{3-9}
& & $k=1$ & $k=3$ & $k=5$ & $k=10$ & $k=5$ & $k=10$ & $k=20$ \\ \hline
\multirow{3}{*}{r/4chan} & Time & 0.0 & 0.0 & 0.13  & 0.513 & 0.56 & 2.799 & 8.43  \\
 & Time+Sentiment & {\bf 0.682} & 0.483 & {\bf 0.585} & 0.793 & 2.256 & 8.62 & 20.728 \\
 & Time+Relevance & 0.544 & 0.451 & 0.576 & 0.806 & 2.681 & 9.294 & 21.128 \\
 & Time+Content & 0.588 & 0.473 & 0.579 & 0.782 & 2.327 & 8.725 & 21.719 \\
 & All & {\bf 0.643} & 0.483 & 0.588 & 0.795 & 2.183 & 8.519 & 20.721 \\
\hline
\multirow{3}{*}{r/AskHistorians} & Time & 0.0 & 0.0 & 0.382 & 0.84 & 0.896 & 2.549 & 4.417 \\
 & Time+Sentiment & {\bf 0.667} & 0.514 & {\bf 0.744} & {\bf 0.922} & 2.306 & 5.396 & 10.396 \\
 & Time+Relevance & 0.437 & 0.431 & {\bf 0.688} & 0.896 & 2.924 & 7.646 & 13.285 \\
 & Time+Content & 0.563 & 0.468 & 0.696 & 0.936 & 2.514 & 7.09 & 10.812 \\
 & All & 0.528 & 0.486 & {\bf 0.708} & {\bf 0.917} & 2.569 & 6.944 & 11.542 \\
\hline
\multirow{3}{*}{r/AskReddit} & Time & 0.0 & 0.0 & 0.181 & 0.678 & 2.258 & 8.553 & 16.249 \\
 & Time+Sentiment & 0.254 & 0.285 & 0.457 & 0.796 & 1.197 & 6.072 & 12.931 \\
 & Time+Relevance & 0.251 & 0.297 & 0.503 & 0.82 & 1.292 & 5.301 & 10.603 \\
 & Time+Content & 0.19 & 0.287 & 0.507 & 0.828 & 1.296 & 5.292 & 11.023 \\
 & All & 0.193 & 0.295 & 0.485 & 0.821 & 1.197 & 5.452 & 11.469 \\
\hline
\multirow{3}{*}{r/askscience} & Time & 0.0 & 0.0 & 0.379 & 0.813 & 1.462 & 4.643 & 9.72 \\
 & Time+Sentiment & {\bf 0.412} & 0.396 & {\bf 0.67} & {\bf 0.888} & 2.099 & 4.758 & 8.176 \\
 & Time+Relevance & 0.253 & 0.385 & 0.664 & 0.897 & 2.198 & 5.198 & 8.533 \\
 & Time+Content & 0.368 & 0.368 & 0.63 & 0.897 & 1.94 & 5.115 & 8.593 \\
 & All & {\bf 0.456} & 0.405 & {\bf 0.664} & {\bf 0.909} & 1.857 & 4.538 & 7.39 \\
\hline
\multirow{3}{*}{r/Bitcoin} & Time & 0.0 & 0.0 & 0.224 & 0.714 & 1.534 & 5.316 & 11.302 \\
 & Time+Sentiment & {\bf 0.465} & 0.411 & {\bf 0.624} & {\bf 0.868} & 1.9 & 6.246 & 11.493 \\
 & Time+Relevance & 0.426 & 39 & 0.583 & 0.864 & 1.839 & 6.617 & 12.444 \\
 & Time+Content & 0.388 & 0.38 & 0.589 & 0.841 & 2.022 & 6.861 & 13.857 \\
 & All & 0.407 & 0.387 & 0.595 & 0.842 & 1.971 & 6.652 & 13.111 \\
\hline
\multirow{3}{*}{r/conspiracy} & Time & 0.0 & 0.0 & 0.23 & 0.662 & 1.221 & 4.312 & 9.257 \\
 & Time+Sentiment & {\bf 0.643} & 0.467 & {\bf 0.638} & {\bf 0.858} & 1.857 & 6.354 & 13.207 \\
 & Time+Relevance & 0.51 & 0.444 & 0.642 & 0.851 & 2.232 & 6.787 & 14.156 \\
 & Time+Content & {\bf 0.618} & 0.462 & {\bf 0.635} & {\bf 0.849} & 1.939 & 6.49 & 13.698 \\
 & All & 0.563 & 0.457 & 0.602 & 0.823 & 1.92 & 6.559 & 14.994 \\
\hline
\multirow{3}{*}{r/news} & Time & 0.0 & 0.0 & 0.162 & 0.47 & 0.922 & 4.044 & 13.371 \\
 & Time+Sentiment & {\bf 0.679} & 0.437 & 0.553 & 0.756 & 1.821 & 6.554 & 17.446 \\
 & Time+Relevance & 0.524 & 0.392 & 0.54 & 0.741 & 2.068 & 7.615 & 19.966 \\
 & Time+Content & 0.561 & 0.398 & 0.531 & 0.718 & 2.003 & 7.098 & 19.22 \\
 & All & 0.541 & 0.401 & 0.514 & 0.703 & 1.993 & 7.233 & 21.405 \\
\hline
\multirow{3}{*}{r/science} & Time & 0.0 & 0.0 & 0.171 & 0.515 & 1.149 & 4.796 & 14.307 \\
 & Time+Sentiment & {\bf 0.502} & 0.372 & {\bf 0.542} & 0.71 & 1.932 & 6.33 & 15.194 \\
 & Time+Relevance & {\bf 0.498} & 0.388 & {\bf 0.522} & 0.693 & 1.9 & 6.476 & 16.388 \\
 & Time+Content & 0.485 & 0.351 & 0.477 & 0.644 & 1.896 & 6.893 & 17.042 \\
 & All & {\bf 0.51} & 0.384 & {\bf 0.515} & 0.69 & 1.919 & 6.497 & 14.506 \\
\hline
\multirow{3}{*}{r/todayilearned} & Time & 0.0 & 0.0 & 0.176 & 0.51 & 0.969 & 4.111 & 12.311 \\
 & Time+Sentiment & {\bf 0.714} & 0.445 & 0.538 & 0.703 & 1.827 & 7.034 & 18.933 \\
 & Time+Relevance & {\bf 0.646} & 0.437 & 0.55 & 0.734 & 2.127 & 7.14 & 18.014 \\
 & Time+Content & {\bf 0.647} & 0.433 & 0.533 & 0.701 & 2.231 & 7.729 & 20.488 \\
 & All & {\bf }0.627 & 0.436 & 0.54 & 0.717 & 2.033 & 7.466 & 20.202 \\
\hline
\multirow{3}{*}{r/worldnews} & Time & 0.0 & 0.0 & 0.153 & 0.462 & 1.05 & 4.045 & 13.259 \\
 & Time+Sentiment & {\bf 0.671} & 0.451 & {\bf 0.558} & 0.705 & 1.912 & 7.183 & 19.847 \\
 & Time+Relevance & {\bf 0.575} & 0.417 & {\bf 0.526} & 0.707 & 1.94 & 7.804 & 20.99 \\
 & Time+Content & {\bf 0.58} & 0.429 & {\bf 0.518} & 0.707 & 2.06 & 8.173 & 21.387 \\
 & All & {\bf 0.601} & 0.425 & {\bf 0.534} & 0.723 & 2.055 & 8.048 & 19.774 \\
\hline
\multirow{3}{*}{r/worldpolitics} & Time & 0.0 & 0.0 & 0.197 & 0.658 & 0.923 & 3.359 & 8.509 \\
 & Time+Sentiment & {\bf 0.722} & 0.519 & {\bf 0.653} & {\bf 0.844} & 2.115 & 6.91 & 15.705 \\
 & Time+Relevance & 0.491 & 0.427 & 0.626 & 0.835 & 2.637 & 8.218 & 17.197 \\
 & Time+Content & 0.573 & 0.503 & 0.64 & 0.857 & 2.209 & 7.551 & 17.949 \\
 & All & 0.615 & 0.489 & 0.625 & 0.842 & 2.06 & 7.594 & 17.487 \\
\hline
\end{tabular}
\end{table*}

\subsection{There are both general and community specific factors that distinguish highly ranked comments.}
\begin{figure*}[ht]
\centering
\includegraphics[width=\textwidth]{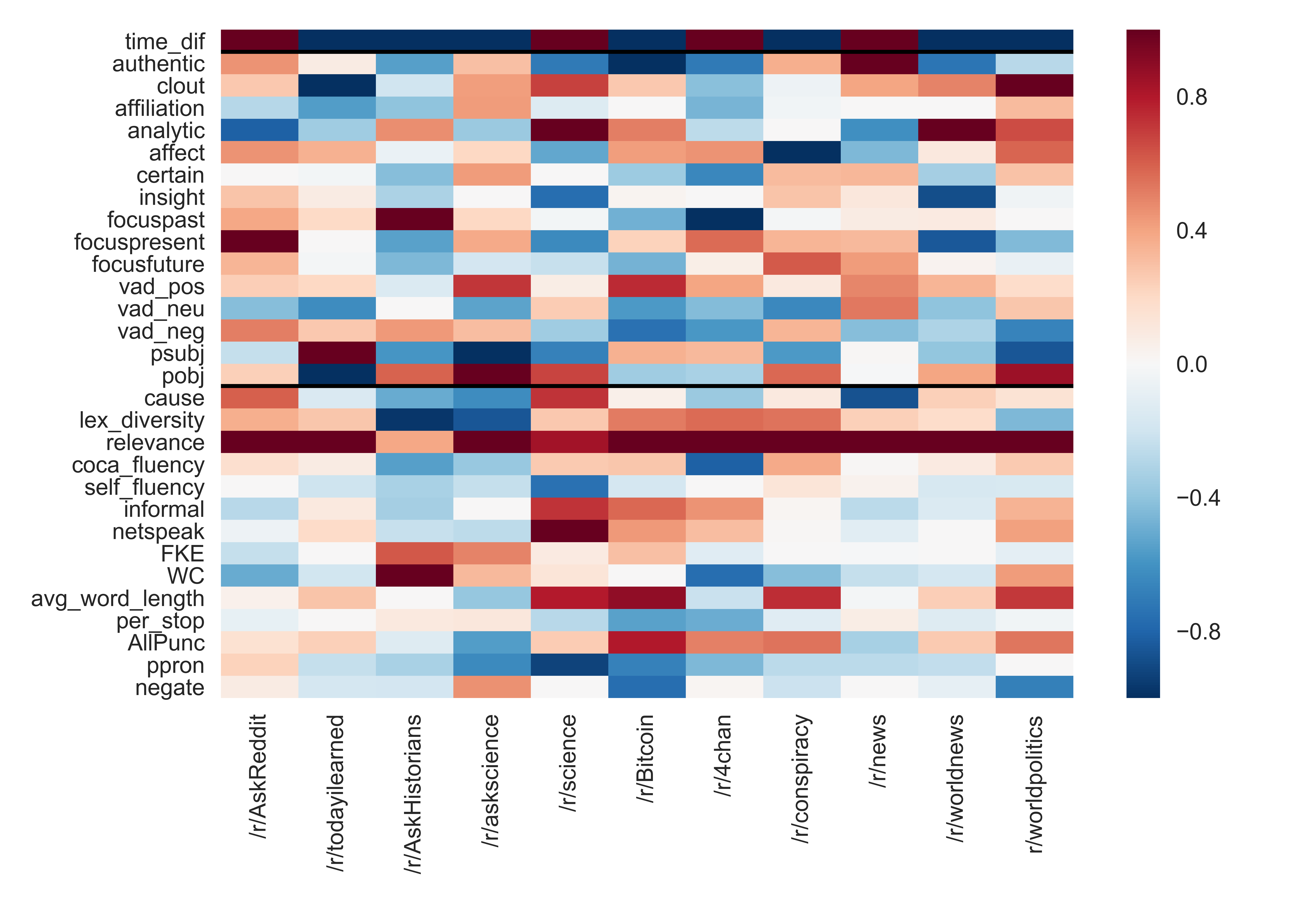}
\caption{\label{fig:heat}A selection of the most important features
  according to the effect size using the KS statistic. For example, high
  scored comments in \texttt{r/AskHistorians,r/askscience} and
  \texttt{r/science} had lower subjectivity (psubj) given by blue and
  higher objectivity (pobj) given by red. The intensities signify their relative effect size among the sentiment features within each subreddit.}
\end{figure*}

To address our second main question, we perform a post-hoc analysis of
the feature distribution for high and low score comments to determine
which features distinguish high score comments, how this changes
across communities, and how this corresponds to the explicit and
implicit rules of the corresponding subreddits. The most important
results can be found in Figure~\ref{fig:heat}. The colors correspond
to whether the feature is positively or negatively impacted the score of
the comment on average while the intensity corresponds to the relative
effect size normalized over the effect sizes of all features for each
subreddit.

\begin{figure*}[ht]
  \centering
  \begin{tabular}{ccc}
    \includegraphics[width=0.3\linewidth]{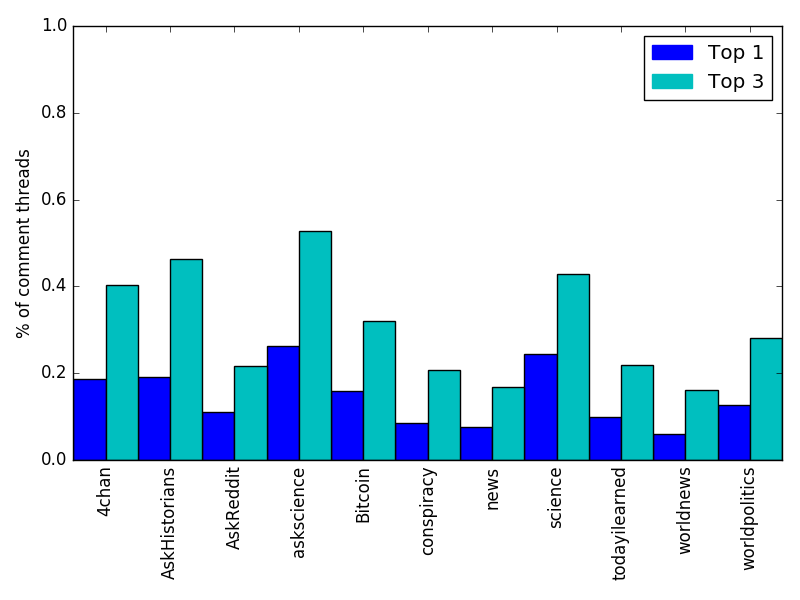} & \includegraphics[width=0.3\linewidth]{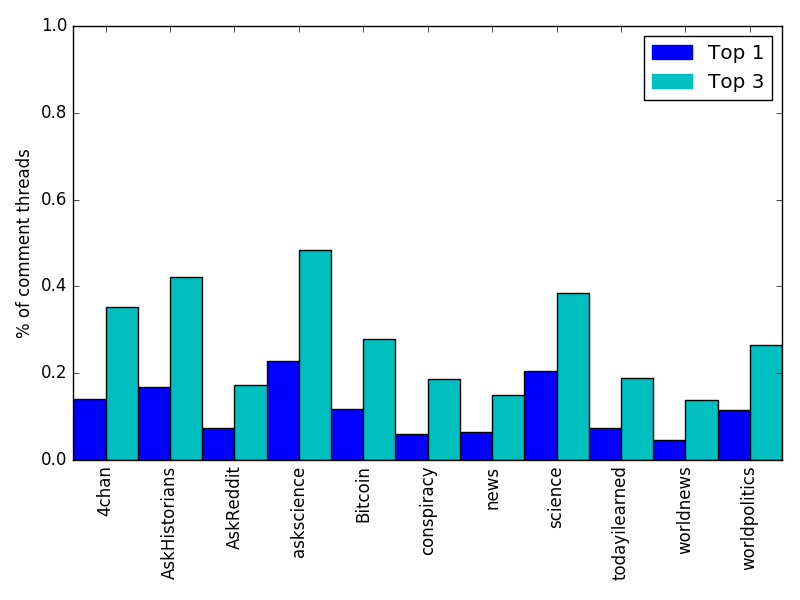} &
    \includegraphics[width=0.3\linewidth]{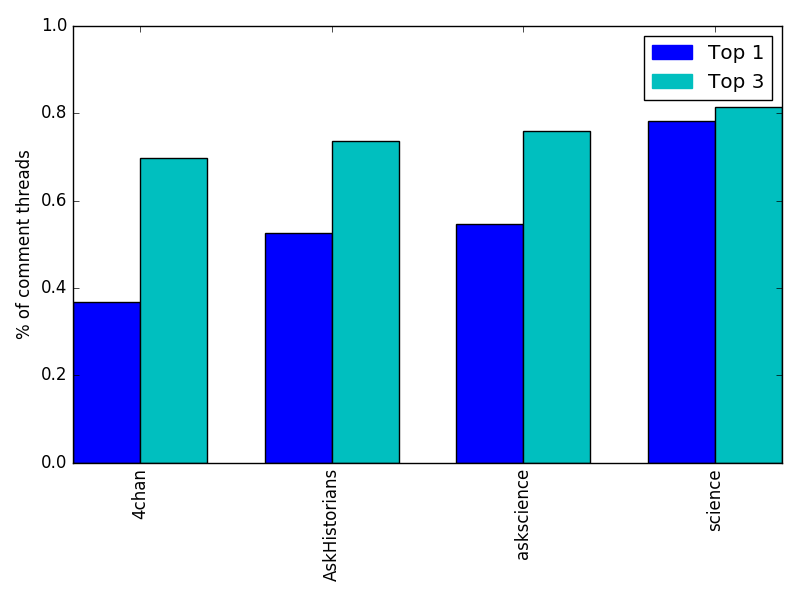} \\
  \end{tabular}
  \caption{Percent of comment threads with top comment by
    (\textbf{left}) the top h-index user (\textbf{center}) the top active
    user, and (\textbf{right}) by a user with a flair.}\label{allusers}
\end{figure*}

\subsubsection{Timeliness is always important, but differently across communities}
Saliently, we find that the importance of comment timing relative to
the post is not consistent across communities. Specifically, we find
that in the communities \texttt{r/AskReddit}, \texttt{r/science},
\texttt{r/4chan}, and \texttt{r/news}, comments that are made later in
time tend to have higher scores, while the rest of the communities
show the opposite effect. Previously, it has been shown that timing
impacts the popularity of posts, in particular the time of the day or
the week~\cite{hessel2017cats}. It has also been shown that comments
which are submitted close the post submission time elicite more
responses in the Multiple Inquirer Single Responder communtity
\texttt{r/IAmA}~\cite{dahiya2016discovering}. Our result shows that
the timing relative to the post is more dependent on the community
than previously thought.

This result may appear for different reasons. For example,
\texttt{r/AskReddit} often has posts reach and stay on the front page
of reddit for a full day or more. This extended time may gain
attention from multiple bursts of people, creating new sets of
comments and new sets of votes later in time. While this bursty
behavior inherently will not change the score of a popular post by
much, it may change the number of comments and votes on comments by
significant amounts. Similarly, timing of comments may be impacted by
the average number of posts submitted to the community.

It is important to note that this does not nessarily negate the
well-known rich-get-richer phenomenon on
reddit~\cite{salganik2006experimental}~\cite{hessel2017cats}, but says
that the rich-get-richer effect may not hold as strongly for comments
as it does for posts.

\subsubsection{Being relevant always matters}
As expected, we find that comments that are more relevant to the post
garner higher scores. This feature is the only feature that is
globally consistent across all communities. Comment relevence was also
shown to be important across several communities
in~\cite{jaech2015talking}.

\subsubsection{New information over stale memes}
Interestingly, we find that writing comments within community
vocabulary (self\_fluency) is
not very important in comment popularity; in fact, it may hurt a
comment's popularity. Specifically, we find that high score comments in
expertise communities have a low self-fluency. This may mean the low
score comments contain memes or jokes that have cycled in the community
before or simply contain old information.

An alternate, and maybe more accurate, interpretation of this feature
is how much new or rare information is in a comment relative to the
community's history. Since this feature is the average frequency of
highly frequent words in a corpus of community text, we are likely
capturing how new or rare the information. This interpretation aligns
well with how we expect expert communities to behave, as new
information is more valuable information.

\subsubsection{Moderation does not always impact behavior}
Contrary to what we expected, moderation has much less of an impact on
the normality of community behavior. In Figure~\ref{fig:heat}, we can
see significant similarity between \texttt{r/worldnews} and
\texttt{r/worldpolitics} across many features. These common features
include preferring more objective comments, less negative comments,
more analytic comments, comments showing clout, and longer average
word length. While the communities cover similar topics
(i.e. international news), they are moderated in explicitly opposite
ways. Figure~\ref{collage}a shows the completely opposing moderation
structure of these two communities:\texttt{r/worldnews} provides well
moderated, non-opinionated news stories, while
\texttt{r/worldpolitics} has no restrictions on what news should be
submitted, explicitly allowing propaganda, fake news, and offensive
content. This notion is further supported by our prediction results,
in which the strictness of moderation does not drastically impact our
ability to rank comments.

Strikingly, these commonalities do not extend to the topic of news in
general, as the community \texttt{r/news} (i.e. U.S. news) is
exceedingly different from \texttt{r/worldnews} despite very similar
moderation restrictions. Particularly, we find that \texttt{r/news}
prefers much less emotional comments, that are more authentic and less
analytic than both \texttt{r/worldnews} and
\texttt{worldpolitics}. Along with this, comments that are made long
after the post submission in \texttt{r/news} may get higher score
comments, while \texttt{r/worldnews} and \texttt{worldpolitics} have
more short lived comment threads.

\subsubsection{Audience can change behavior more than topic}
The generality of a community's audience can also steer the community
discussion. For example, we can see a distinct divide between the
expert communities based on how niche the target audience is.  The
communities \texttt{r/AskHistorians} and \texttt{r/askscience} are
built for explaining questions to a wide audience, while
\texttt{r/science} and \texttt{r/Bitcoin} are build for discussions
among a narrow audience (scientists and Bitcoin experts). Intuitively,
\texttt{r/AskHistorians} and \texttt{r/askscience} prefer comments
that are more lexically redundant, have less causation words (i.e
think, know), and use more fluent or common terms. All of these
features reflect the behavior of simply explaining the answer to a
question. On the other hand, we see \texttt{r/science} and
\texttt{r/Bitcoin} prefer less lexically redundant comments that use
more technical and analytic terms.

\subsubsection{Even communities of the same type differ}
It is clear from our results, and from the
literature~\cite{jaech2015talking} that online communities differ
vastly in discussion style and conduct. While we find many of the same
features important in all communities, their direction of importance
may change. For example, the authenticity of a comment is important
across all subreddits; however, some prefer more authenticity, others
significantly less. In general, emotion is important across all
subreddits, but some significantly dislike emotional comments, where
as others significantly prefer emotional comments.  Even very similar
communities, like \texttt{r/worldnews} and \texttt{r/worldpolitics},
small differences can be found. \texttt{r/worldpolitics} uses
significantly more netspeak and informal words, while
\texttt{r/worldnews} uses more lexical diversity in discussion.

\subsubsection{Highly active and high scoring users do not have a significant effect in the reception of comments, but existence of flairs do}
We compute the percent of comment threads in which the top 1 and top 3
comments is by a user with the highest h-index, the highest activity,
or has a flair as shown in Figure~\ref{allusers}. We find that both
the local reputation and the local activity levels of a user have
little impact on the score of the comment. This result is consistent
with the literature~\cite{jaech2015talking}. More significantly, we
find that users who have a flair have a high chance of being the
highest score in a comment thread, especially if those flairs are
expert flairs, despite only 1\% to 5\% of users having flairs (see
Figure~\ref{subs}). Further, we find that flaired users are more
active overall and have a higher local h-index overall. In terms of
expert flairs, this result is consistent with what we know about
experts online~\cite{horne2016expertise}. We also test whether some
flairs are more important than others, but find nothing significant.

\section{Conclusion and Future Work}
We confirm that it is possible to predict the score of comments, even when communities are unstructured, loosely moderated and noisy. Leveraging a carefully crafted set of features, our models outperform the state of the art model by a factor of 2, achieving high average precision and show that the relative rankings of comments remains close to the true ranking. Interestingly, the importance of features can vary vastly across communities. Despite this, we show some globally important features, such as relevance and emotion
of comments and discuss potential reasons for the differences. Further, we show that user flairs can be an excellent
predictor of highly popular comments, especially if those flairs are
strictly controlled by moderators.


In the future, we would like to study what impact the users have on
discussion, by conducting time-controlled experiments and
collecting more user specific data such as years on reddit or if the
user is a moderator of a community. Along with this, we would like to
take a deeper look at the impact of different levels of
moderation. While we show compelling results on the surprisingly small
impact strict moderation has on discussion behavior, there may be many
other behaviors impacted by moderation.
We also would like to group different subreddits based on their
similarities across our features. Another direction is to study how
our findings change over time to see if changes to reddit scoring
methods and other world events impact the findings.

\section{Acknowledgments}
{\footnotesize Research was sponsored by the Army Research Laboratory and was
accomplished under Cooperative Agreement Number W911NF-09-2-0053 (the
ARL Network Science CTA). The views and conclusions contained in this
document are those of the authors and should not be interpreted as
representing the official policies, either expressed or implied, of
the Army Research Laboratory or the U.S. Government. The
U.S. Government is authorized to reproduce and distribute reprints for
Government purposes notwithstanding any copyright notation here on.}
\bibliographystyle{IEEEtran}
\bibliography{IEEEabrv,references}

\end{document}